\documentclass[12pt]{article}
\usepackage{amsmath}
\def\be{\begin{equation}}
\def\ee{\end{equation}}
\def\arr{\begin{array}{rll}}
\def\ea{\end{array}}
\def\bea{\begin{eqnarray}}
\def\eea{\end{eqnarray}}

\def\N2{$N{=}2$}

\def\tr{{\rm tr}}

\def\>{\rangle}
\def\<{\langle}
\def\+{\dagger}
\def\={\ =\ }

\begin{document}
\renewcommand{\thefootnote}{\fnsymbol{footnote}}
\begin{titlepage}
\setcounter{page}{0}
\begin{center}
{\LARGE\bf  The Bohlin variant of the Eisenhart lift}\\
\vskip 1.2cm
\textrm{\Large Anton Galajinsky \ }
\vskip 0.7cm
{\it
Tomsk Polytechnic University, 634050 Tomsk, Lenin Ave. 30, Russia} \\
\vskip 0.2cm
{e-mail: galajin@tpu.ru}
\vskip 0.5cm
\end{center}
\vskip 1cm
\begin{abstract} \noindent
Inspired by the Bohlin transformation relating the planar harmonic oscillator to the Kepler problem, a variant of the
Eisenhart lift is studied, in which a Lagrangian conservative dynamical system with $d$ degrees of freedom is embedded
into timelike geodesics of a conformally flat metric on a $(d+2)$--dimensional space--time of the Lorentzian signature. 
The uplift is used to construct novel examples of conformally flat metrics admitting higher rank Killing tensors.
\end{abstract}

\vspace{0.5cm}

\noindent

\end{titlepage}

\renewcommand{\thefootnote}{\arabic{footnote}}
\setcounter{footnote}0

\noindent
{\bf 1. Introduction}\\

Since the formulation of general relativity, geometry
and theoretical physics go parallel and mutually enriching courses. In particular, 
geometrization of a Lagrangian
conservative mechanics amounts to embedding
Newton's equation into the null geodesics of the Eisenhart metric \cite{LE,DBKP}
\be
\mathcal{L}(x,\dot x) dt^2+dt ds,
\nonumber
\ee
where $\mathcal{L}(x,\dot x)$ is the Lagrangian density, $t$ is the temporal coordinate, and
$s$ is an extra variable \cite{LE,DBKP}. The latter is needed in order to reproduce the conventional
nonrelativistic energy conservation condition within the general relativistic framework. At the level of 
equations of motion, $s$ links
to the action variable sometimes adopted in classical 
mechanics. Thus, given
a dynamical system with $d$ degrees of freedom $x_i$, $i=1,\dots,d$, the 
resulting space--time
is $(d+2)$--dimensional and the metric is of the Lorentzian signature.

A salient feature
of Eisenhart's metric is that it admits a covariantly constant null Killing vector field
$\frac{\partial}{\partial s}$, meaning that the space--time belongs to the Kundt class,
as it contains 
a geodesic null congruence with vanishing expansion, shear and vorticity. 

If a Lagrangian density explicitly depends on time, $\mathcal{L}(t,x,\dot x)$, the metric is referred to as the 
Brinkmann--type metric or a plane--fronted wave with parallel rays 
(the $pp$--wave).

As is well known,
in Riemannian geometry symmetries of a metric are described by Killing vectors, which 
generate infinitesimal coordinate
transformations in space--time, $y'^A=y^A+\xi^A(y)$. 
As far as the geodesic equations are concerned, each 
Killing vector field with components $\xi^A(y)$
gives rise to the first integral $\xi^A(y) g_{AB} (y) \frac{d y^A}{d \tau}$ of the geodesic equations, 
where $g_{AB} (y)$ is a metric and $\tau$ is the proper time. 

A space--time may also admit Killing tensors, 
i.e. totally symmetric tensor fields obeying the equation $\nabla_{(A_1} K_{A_2 \dots A_{n+1} ) }(y)=0$, 
which underlie first integrals of the form 
$K_{A_1 \dots A_n} (y) \frac{d y^{A_1}}{d \tau}$ \\
$\dots \frac{d y^{A_n}}{d \tau}$. 
The existence of Killing tensors is usually attributed to hidden symmetries of a
space--time, as 
there is no coordinate transformation associated to them.

The importance of the Killing vectors and tensors is hard to overestimate. They give a clue to
establishing complete integrability of the geodesic equations formulated in various black
hole space--times, as well as allow one to separate variables in the Hamilton–Jacobi, 
Klein--Gordon and Dirac equations in strong gravitational fields (for a review and further references see \cite{FKK}).

Apart from the aesthetic appeal, the Eisenhart lift
provides an efficient means of constructing novel space--times with hidden symmetries. By 
applying the method to various integrable models, one can build a plethora of examples of such a kind
\cite{DGH}--\cite{BK}. Some of them provide exact solutions to the Einstein equations
\cite{DGH,CG,GF,FG}. Worth mentioning also is a
considerable body of recent work on application of the Eisenhart lift to cosmology (see e.g.
\cite{G1}--\cite{AP} and references therein).

The goal of this paper is to study a variant of the Eisenhart lift, which is inspired by the Bohlin
transformation \cite{B} (for a review see \cite{S})
connecting the planar harmonic oscillator to the Kepler problem. Given a dynamical system 
with $d$ degrees of freedom, a simple conformally flat metric on a $(d+2)$--dimensional space--time 
of the Lorentzian signature is proposed, timelike geodesics of which include the original Newton equation.

The work is organized as follows. 

In the next section, we briefly outline the Eisenhart lift, 
paying particular attention to physical dimensions of the ingredients involved. 

In Sect. 3, 
by formally applying the Bohlin
transformation to the Eisenhart metric originating from the two--dimensional 
harmonic oscillator, we obtain a simple conformally flat metric (the conformal 
factor being the Kepler potential), 
which can be readily
generalized to the case of arbitrary dimension and potential. In contrast to the Eisenhart framework,
the metric does not belong to the Kundt class and the original Newton equation 
is recovered by analyzing timelike geodesics rather than null geodesics. The issue of hidden symmetries is discussed as well.

Sect. 4 contains explicit examples.
Firstly, it is shown that the celebrated conformal mechanics of \cite{AFF}, when Bohlin uplifted, yields
a metric of the three--dimensional anti de Sitter space. Secondly, the four--body Calogero model \cite{Cal1} 
is used to build a six--dimensional conformally flat metric admitting irreducible 
Killing tensors of rank 3 and 4. By increasing the number of particles 
in the original Calogero model, one can build
a $(d+2)$--dimensional Bohlin--type metric admitting Killing tensors 
of rank up to $d$. 

We summarize our results and discuss possible further developments in the concluding Sect. 5.

Throughout the paper, summation over repeated indices is understood unless otherwise stated.

\vspace{0.5cm}

\noindent
{\bf 2. The Eisenhart lift}\\

Consider a conservative dynamical system with $d$ degrees of freedom $x_i$, $i=1,\dots,d$,
and let $t$ be the temporal variable. Assuming that
potential energy $V(x)$ can be factorized $V(x)=g U(x)$, where $g$ is 
a constant having the dimension of energy and $U(x)$ is a positive dimensionless
function,\footnote{Because the coordinates $x_i$ have the dimension of length and 
$U(x)$ is assumed to be dimensionless, the latter will necessarily involve a (coupling) constant
with the dimension of length. By this reason, a numerical value of $g$ entering 
the original potential $V(x)=g U(x)$ is of no great importance. In what follows, we refer to $U$ as the prepotential.} the Newton equation
\be\label{NE}
m \frac{d^2 x_i}{d t^2}+\frac{\partial V(x)}{\partial x_i}=0,
\ee
where $m$ is particle's mass, can be embedded into the null geodesics 
\be\label{geo}
\frac{d^2 y^A}{d \lambda^2}+\Gamma^A_{BC}  \frac{d y^B}{d \lambda} 
\frac{d y^C}{d \lambda}=0, 
\qquad g_{AB} \frac{d y^A}{d \lambda} \frac{d y^B}{d \lambda}=0
\ee
of the Eisenhart metric \cite{LE,DBKP}
\be\label{metric}
g_{AB} d y^A d y^B=2 U(x) c^2 d t^2-2 c dt ds-d x_i dx_i,
\ee
where $U(x)$ is the prepotential above, $s$ is an extra coordinate having the dimension of length, 
$c$ is the speed of light, 
$y^A=(c t,s,x_1,\dots,x_d)$, and $\lambda$ is an affine parameter (of the dimension of time) along a null geodesic. 

Indeed, computing the Christoffel symbols
\be\label{cs}
\Gamma^i_{tt}=-\Gamma^s_{t i}=\partial_i U,
\ee
where the notation $A=(t,s,i)$, $\partial_i=\frac{\partial}{\partial x_i}$,
with $i=1,\dots,d$, was adopted,
one can reduce the first equation in (\ref{geo}) to
\be\label{seom}
\frac{d t}{d \lambda}=\alpha, \qquad 
\frac{d s}{d \lambda}-2 \alpha c U=c \beta, 
\qquad m \frac{d^2 x_i}{d t^2}+m c^2 \partial_i U=0,
\ee
where $\alpha>0$ (forward in time motion) and $\beta$ are dimensionless real constants of integration. 
Note that in the last equation $\frac{d}{d\lambda}=\frac{d t}{d\lambda}\frac{d}{dt}$ 
was used and the common factor ${\left(\frac{d t}{d \lambda} \right)}^2$ was dropped.

The second equation in (\ref{geo}) relates the ratio $\frac{\beta}{\alpha}$ to
particle's energy 
\be\label{ener}
\frac{m}{2}  \frac{d x_i}{dt} \frac{d x_i}{dt}+m c^2 U(x)=-\frac{m c^2 \beta}{\alpha},
\ee
as well as allows one to determine how the extra variable $s$ evolves with time
\be\label{efs}
s(t)=-\frac{1}{mc} \int_0^t d \tilde t 
\left(\frac{m}{2} \frac{d x_i \left(\tilde t \right)}{d \tilde t} 
\frac{d x_i \left(\tilde t \right)}{d \tilde t}
-m c^2 U\left( x \left(\tilde t \right) \right) \right)+s_0,
\ee
where $s_0$ is a constant of integration. The latter formula implies that 
$s$ links to the action variable sometimes adopted in classical 
mechanics.\footnote{In principle, one is allowed 
to consider 
timelike geodesics within the Eisenhart framework. In that case,
the right hand side of (\ref{efs}) would involve an extra contribution 
$-\frac{c t}{2 \alpha^2}$, 
which would
only spoil the interpretation of $s(t)$ as the action variable. 
A numerical value of the energy in (\ref{ener}) would also
include 
$-\frac{m c^2}{2\alpha^2}$, thus yielding 
$-\frac{m c^2 (1+2 \alpha \beta)}{2 \alpha^2}$ on the right hand side of (\ref{ener}).}

The Newton equation (\ref{NE}) is thus recovered by identifying $m c^2 U(x)=V(x)$ and
implementing the null reduction along $s$. Note that in physics literature it is customary to use units in which $c=1$ and 
the mass of a particle is usually set to unity. 
 
Let us briefly discuss properties of the Eisenhart metric (\ref{metric}). Computing the eigenvalues of $g_{AB}$
\be
U+\sqrt{U^2+1}, \qquad U-\sqrt{U^2+1}, \qquad -1, \qquad \dots \qquad, -1,
\ee
one concludes that (\ref{metric}) describes a $(d+2)$--dimensional space--time of 
the Lorentzian signature. 

The metric admits two Killing vector fields
\be\label{KV}
\xi^A \frac{\partial}{\partial y^A}=\frac{\partial}{\partial t}, \qquad
\eta^A \frac{\partial}{\partial y^A}=\frac{\partial}{\partial s}.
\ee
The corresponding first integrals of the geodesic equations 
\be
\xi^A g_{AB} \frac{d y^B}{d \lambda}, \qquad
\eta^A g_{AB} \frac{d y^B}{d \lambda}
\nonumber
\ee
link to the first two equations in (\ref{seom}). In general relativity, a 
space--time is called stationary 
if it admits a timelike Killing vector field. For the case at hand, $\xi^A$ is timelike
provided the prepotential is a positive function $U(x)>0$. The second Killing vector 
$\eta^A$ is null and, more importantly, it is covariantly constant $\nabla_A \eta_B=0$.
Hence, (\ref{metric}) admits
a geodesic null congruence with vanishing expansion, shear and vorticity 
and thus belongs to the Kundt class.

Computing the Ricci tensor, one reveals the only
non--vanishing component 
\be
R_{tt}=\Delta U,
\nonumber
\ee
which also means that the scalar curvature vanishes, harmonic prepotentials 
uplift to Ricci--flat metrics, and no cosmological term is allowed.

Finally, assuming the original dynamical system (\ref{NE}) admits a 
polynomial integral of motion of degree $n$ in velocities
\be\label{FI}
\mathcal{I}=
\sum_{k=0}^n I_{i_1 \dots i_k} (x) \frac{d x_{i_1}}{dt} \dots 
\frac{d x_{i_k}}{dt}, \qquad \frac{d \mathcal{I}}{d t}=0,
\ee
one can uplift it to a Killing tensor\footnote{For a discussion of 
conformal Killing tensors within the Eisenhart framework see e.g.
\cite{G,CGHHZ}.} of rank $n$. Recall that the geodesic equations (\ref{geo}), (\ref{seom})
imply that the coordinate time $t$ and the affine parameter $\lambda$ are affinely related $\frac{d t}{d \lambda}=\alpha$.
By
multiplying the conserved charge $\mathcal{I}$ with the constant $\alpha^n={\left(\frac{dt}{d \lambda} \right)}^n$, 
where $n$ is the
highest power of the velocity in (\ref{FI}), one can bring the resulting expression to the homogeneous form 
\be
K_{A_1 \dots A_n} (y) \frac{d y^{A_1}}{d \lambda} 
\dots \frac{d y^{A_n}}{d \lambda},
\nonumber
\ee
where $y^A=(c t,s,x_1,\dots,x_d)$,
from which the components of the Killing tensor obeying
$\nabla_{(A_1} K_{A_2 \dots A_{n+1} )}=0$ are read off.

As was mentioned in the Introduction, within the context of general relativity 
Killing tensors describe hidden symmetries of a space--time. The Eisenhart lift thus
provides a convenient framework for building metrics admitting 
higher rank Killing tensors \cite{DGH}--\cite{FG},
some of which are Ricci--flat \cite{DGH,CG,GF,FG}.

\vspace{0.5cm}

\noindent
{\bf 3. The Bohlin variant of the Eisenhart lift}\\

The Bohlin transformation \cite{B} (for a review see \cite{S}) links the planar harmonic oscillator
described by a complex coordinate $w$ and the temporal variable $t$ to the Kepler problem 
specified by a complex coordinate $z$, the evolution parameter $\lambda$,
and a negative value of conserved energy
\be\label{BT}
z=w^2, \qquad w \bar{w} \frac{dt}{d \lambda}=1.
\ee
Note that it is the potential of the original dynamical system (evaluated on particle's orbit) which determines the new 
temporal coordinate $\lambda$. For simplicity of the presentation, in (\ref{BT})
we ignored physical dimensions.

Building the Eisenhart metric associated with the two--dimensional 
harmonic oscillator and 
formally\footnote{The second equation in (\ref{BT}) involves the general solution $w(t)$ to the 
oscillator equation of motion, meaning that (\ref{BT}) cannot be regarded as a proper coordinate transformation
within the context of general relativity.} implementing (\ref{BT}) there, one obtains a 
simple conformally flat metric, the conformal factor being the Kepler potential, 
which can be readily
generalized to the case of arbitrary dimension and prepotential
\be\label{BM}
g_{AB}(y) d y^A d y^B=U(x) \left(2 c dt ds-d x_i dx_i \right),
\ee
where $y^A=(c t,s,x_1,\dots,x_d)$, $c$ is the speed of light, and $U(x)$ is an arbitrary positive function. 

Let us discuss properties of (\ref{BM}). The eigenvalues of $g_{AB}$ 
are $\left(U,-U,\dots,-U \right)$ and, hence,
it describes a $(d+2)$--dimensional conformally flat stationary space--time 
of the Lorentzian signature. 

For an arbitrary $U(x)$ it admits three Killing vector fields 
\be\label{KV1}
\frac{\partial}{\partial t}, \qquad
\frac{\partial}{\partial s}, \qquad t \frac{\partial}{\partial t}-s \frac{\partial}{\partial s},
\ee
which form Lie algebra of the Poincar\'e group in $1+1$ dimensions. 
None of them is covariantly constant and, hence, the space--time does not belong to
the Kundt class.
If $U(x)$ is a homogeneous function of degree $-2$, the metric holds invariant under the 
dilatation transformation $y'^A=\sigma y^A$, where $\sigma$ is a constant. Such metrics might be of potential interest for holographic applications.

Computing the Christoffel symbols
\bea\label{CS}
&&
\Gamma^t_{ti}=\Gamma^s_{s i}=\Gamma^i_{ts}=\frac 12 U^{-1} \partial_i U, \qquad
\Gamma^i_{jk}=\frac 12 U^{-1} \left(\partial_j U \delta_{ik}+
\partial_k U \delta_{ij}-\partial_i U
\delta_{jk} \right), 
\nonumber\\[2pt]
&&
\Gamma^A_{A i}=\frac{(d+2)}{2} U^{-1} \partial_i U,
\eea
where $\delta_{ij}$ is the Kronecker delta and $\partial_i=\frac{\partial}{\partial x_i}$, one can then determine the Ricci tensor
\bea
&&
R_{ts}=\frac{1}{2} U^{-1} \left(\Delta U+\frac{(d-2)}{2} U^{-1} \partial_i U \partial_i U \right),
\nonumber\\[2pt]
&&
R_{ij}=-\frac{d}{2} U^{-1} \left( \partial_i \partial_j U
-\frac{3}{2} U^{-1} \partial_i U \partial_j U\right)-R_{ts} \delta_{ij}
\eea
and the scalar curvature
\be
R=(d+1) U^{-2} \left(\Delta U+\frac{(d-4)}{4} U^{-1} \partial_i U \partial_i U \right).
\ee
Taking into account the identity
\be
\partial_i \partial_j U^k=k U^{k-1} \left(\partial_i \partial_j U+(k-1) 
U^{-1} \partial_i U \partial_j U \right),
\ee
one concludes that the space--time is Ricci--flat only for a trivial prepotential 
$U=\mbox{const}$, while the scalar curvature vanishes for
\be
U=F^{\frac{4}{d}}, \qquad \Delta F=0.
\ee 

Now let us turn to the timelike geodesics 
\be\label{Bg}
\frac{d^2 y^A}{d \tau^2}+\Gamma^A_{BC} \frac{d y^B}{d \tau} 
\frac{d y^C}{d \tau}=0, 
\qquad g_{AB} \frac{d y^A}{d \tau} \frac{d y^B}{d \tau}=c^2,
\ee
where $\tau$ is the proper time, which follow from (\ref{BM}). Taking into account (\ref{CS}), the equations
(\ref{Bg}) can be reduced to
\bea\label{BEq}
&&
U \frac{dt}{d\tau}=\alpha, \qquad \frac{1}{c} U \frac{ds}{d\tau}=\beta, \qquad 
m \frac{d^2 x_i}{d t^2}+\frac{m c^2}{2\alpha^2} \partial_i U=0,
\eea
where $\alpha>0$ (forward in time motion) and $\beta$ are dimensionless real constants of integration, 
while the ratio $\frac{\beta}{\alpha}$ links to
particle's energy 
\be
\frac{m}{2} \frac{d x_i}{dt} \frac{d x_i}{dt}+\frac{m c^2}{2 \alpha^2} U(x)=\frac{m c^2 \beta}{\alpha}.
\ee
In contrast to the Eisenhart framework, the dynamics of the extra variable 
$s$ is affinely related to $c t$
\be
\frac{ds}{dt}=\frac{c \beta}{\alpha}.
\ee
Identifying $\frac{m c^2}{2 \alpha^2} U(x)=V(x)$ and implementing the null 
reduction along $s$, one recovers Newton's equation (\ref{NE}).

Although Bohlin's work was published eighteen years prior to Eisenhart's paper, 
it seems tempting to call 
(\ref{BM}) the Bohlin variant of the Eisenhart lift,
because the coordinate time and the proper time are related by the prepotential 
similarly to the second equation 
in (\ref{BT}).

Note that, if (\ref{FI}) happens to be a polynomial integral of motion of a dynamical system at hand,
one can turn it into a Killing tensor of the Bohlin metric. The uplift is less straightforward than 
within the Eisenhart framework, however. Because Newton's equation (\ref{NE}) 
holds invariant under the time reversal transformation $t \to - t$, 
the corresponding first integrals involve monomials in momenta which are all of even/odd power. 
Substituting $p_i=\frac{d x_i}{dt}=\frac{U}{\alpha}\frac{d x_i}{d\tau}$
into an integral of motion of degree $n$ in momenta and multiplying, where necessary, with the unity 
$1={\left(\frac{1}{c^2} g_{AB} \frac{dy^A}{d\tau} \frac{dy^B}{d\tau} \right)}^m$, where $m$ is a natural number,
one can turn the latter into
a homogeneous function $K_{A_1 \dots A_n} \frac{d y^{A_1}}{d \tau} 
\dots \frac{d y^{A_n}}{d \tau}$ of degree $n$ in the velocities $\frac{d y^{A}}{d \tau}$,
from which components of the Killing tensor $K_{A_1 \dots A_n}$ are read off. In contrast to the
conventional Eisenhart lift, higher rank Killing tensors of the Bohlin--type metric will always involve $ts$--components.

The Bohlin lift thus can be used to build novel conformally flat metrics 
of the Lorentzian signature
admitting higher rank Killing tensors. Examples of such a kind are given in the next section.

Concluding this section, we note that a possibility to accommodate 
the Bohlin transformation within the Jacobi geometrization of classical dynamics 
was studied in \cite{CGG}.

\vspace{0.5cm}

\noindent
{\bf 4. Examples}\\

Our first example originates from the prepotential
\be\label{pot}
U(x_1,\dots,x_d)=\frac{1}{{\left(a+b_i x_i \right)}^2}, 
\ee
where $a$ is a dimensionless constant and $b_i$ are constants having the dimension 
of inverse length. When Bohlin uplifted, it results in a metric 
solving the vacuum Einstein equations in the presence of a cosmological 
constant\footnote{As a matter of fact,
(\ref{pot}) yields a metric defined on a patch of the $(d+2)$--dimensional anti de 
Sitter space--time, because
$R_{ABCD}=(b_i b_i) \left(g_{AC} g_{BD}-g_{AD} g_{BC} \right)$. Anti de Sitter and de Sitter space--times thus fit nicely into the framework.}
\be
R_{AB}-\frac 12 R g_{AB}=\Lambda g_{AB}, 
\nonumber
\ee
with $R=(d+1)(d+2)b_i b_i$ and $\Lambda=-\frac{d(d+1) b_i b_i}{2}$.  In particular, 
the celebrated conformal mechanics \cite{AFF}, which
is governed by the potential $V=\frac{g^2}{x^2}$, $g$ being a coupling constant,
belongs to this class. As far as the metric is concerned, the constant $a$ in (\ref{pot})
can be removed by redefining $x_i$. The resulting line element is
invariant under the extra dilatation transformation $y'^A=\sigma y^A$, 
where $\sigma$ is a constant.

Our second example also belongs to the class of scaling invariant metrics and builds upon 
the Calogero model \cite{Cal1}, which describes a set of $d$ identical 
particles on the real line interacting via the inverse--square potential
\be\label{CP}
V(x_1,\dots,x_d)=\frac 12 \sum_{i,j=1,i\ne j}^d \frac{g^2}{{(x_i-x_j)}^2},
\ee
where $g$ is a coupling constant. The model is integrable and the corresponding
first integrals can be obtained from the Lax matrix (no sum over repeated indices below)
\be
L_{ij}=p_i \delta_{ij}+{\rm i} g \left(1-\delta_{ij} \right) {\left(x_i-x_j \right)}^{-1},
\nonumber
\ee
where $p_i (t)=\frac{d x_i (t)}{dt}$ are momenta canonically conjugate 
to coordinates, $\{x_i,p_j\}=\delta_{ij}$, and ${\rm i} $
is the imaginary unit.
Constants of motion are polynomial in the momenta, 
$I_i=\frac{1}{i!} \tr L^i$, with $i=1,\dots,d$. For example, 
the first four members in the string read
\bea\label{intm}
&&
I_1=\sum_{i=1}^d p_i, \qquad
I_2=\frac{1}{2} \sum_{i=1}^d p_i^2  +\sum_{i,j=1,i<j}^d \frac{g^2}{{(x_i-x_j)}^2},
\nonumber\\[2pt]
&&
I_3=\frac{1}{3!} \left(\sum_{i=1}^d p_i^3+3 g^2\sum_{i,j=1,i<j}^d \frac{p_i+p_j}{{(x_i-x_j)}^2}\right),
\nonumber\\[2pt]
&&
I_4=\frac{1}{4!} \left(\sum_{i=1}^d p_i^4+4 g^2\sum_{i,j=1,i<j}^d \frac{p_i^2+p_j^2+p_i p_j}{{(x_i-x_j)}^2}+
2 g^4
\sum_{i,j=1,i<j}^d \frac{1}{{(x_i-x_j)}^4}+
\right.
\nonumber\\[2pt]
&&
\qquad \qquad
\left.
+4 g^4 \sum_{i,j,k=1,i \neq j, i \neq k, j<k}^d \frac{1}{{(x_i-x_j)}^2 {(x_i-x_k)}^2}\right).
\eea

Let us focus on the four--body case and construct a six--dimensional Bohlin--type metric, 
which admits irreducible Killing tensors of rank 3 and 4. For simplicity of the presentation, below we set $m=c=1$.

The prepotential which specifies the metric in (\ref{BM}) reads
\be\label{PP}
U(x_1,\dots,x_4)=\frac 12 \sum_{i,j,i\ne j}^4 \frac{L^2}{{(x_i-x_j)}^2},
\nonumber
\ee
where $L$ is a constant having the dimension of length, which links to the coupling constant 
$g$ in (\ref{CP}) via $g^2=\frac{L^2}{2 \alpha^2}$.
As above, $\alpha>0$ is a constant entering the leftmost relation in (\ref{BEq}). 

Substituting $p_i=\frac{d x_i}{dt}=\frac{U}{\alpha}\frac{d x_i}{d\tau}$
into $I_3$ and $I_4$ in (\ref{intm}) and multiplying the terms involving $g^2$ by the unity 
$1=g_{AB} \frac{dy^A}{d\tau} \frac{dy^B}{d\tau}$ and those containing $g^4$ by 
$1={\left(g_{AB} \frac{dy^A}{d\tau} \frac{dy^B}{d\tau} \right)}^2$, so as to
obtain expressions homogeneous in $\frac{dy^A}{d\tau}$, one finally 
obtains the third--rank Killing 
tensor (no sum over repeated indices unless otherwise stated)
\bea
&&
K_{tsi}=\frac 12 L^2 U^2 \sum_{a=1,a\ne i}^4 \frac{1}{{(x_a-x_i)}^2},
\nonumber\\[2pt]
&&
K_{ijp}=U^3 \delta_{ip} \delta_{jp}-\frac 12 L^2 U^2 
\left(\delta_{ij} \sum_{a=1,a\ne p}^4 \frac{1}{{(x_a-x_p)}^2} 
+\delta_{jp} \sum_{a=1,a\ne i}^4 \frac{1}{{(x_a-x_i)}^2}
\right.
\nonumber\\[2pt]
&&
\left.
\qquad \quad
+\delta_{pi} \sum_{a=1,a\ne j}^4 \frac{1}{{(x_a-x_j)}^2}\right),
\nonumber
\eea
as well as 
the fourth--rank Killing 
tensor 
\bea
&&
K_{ttss}=\frac 16 L^4 U^2 \left(\sum_{a,b=1,a\ne b}^4 \frac{1}{{(x_a-x_b)}^4}+2
\sum_{a,b,c=1,a\ne b \ne c}^4 \frac{1}{{(x_a-x_b)}^2 {(x_a-x_c)}^2}\right),
\nonumber\\[2pt]
&&
K_{tsij}=-\frac{1}{12} L^4 U^2 \delta_{ij} \left(\sum_{a,b=1,a\ne b}^4 \frac{1}{{(x_a-x_b)}^4}+2
\sum_{a,b,c=1,a\ne b \ne c}^4 \frac{1}{{(x_a-x_b)}^2 {(x_a-x_c)}^2}\right) 
\nonumber\\[2pt]
&&
\qquad \qquad
+\frac 16 L^2 U^3 \sum_{a,b=1,a\ne b}^4 \frac{\left(2 \delta_{ai} \delta_{aj}
+\delta_{ai} \delta_{bj}\right)}{{(x_a-x_b)}^2} ,
\nonumber\\[2pt]
&&
K_{ijkp}=U^4 \sum_{a=1}^4 \delta_{ai}  \delta_{aj}  \delta_{ak}  \delta_{ap}
\nonumber\\[2pt]
&& 
+
\frac{1}{12} L^4 U^2 \left( \delta_{ij} \delta_{kp}+ \delta_{ik} \delta_{pj}+
 \delta_{ip} \delta_{jk} \right) \left(\sum_{a,b=1,a\ne b}^4 \frac{1}{{(x_a-x_b)}^4}
+2
\sum_{a,b,c=1,a\ne b \ne c}^4 \frac{1}{{(x_a-x_b)}^2 {(x_a-x_c)}^2}\right) 
\nonumber\\[2pt]
&&
-\frac 16 L^2 U^3 \left(2 \sum_{a,b=1,a \ne b}^4 \frac{\left(
\delta_{ai} \delta_{aj} \delta_{kp}
+\delta_{ai} \delta_{ak} \delta_{pj}
+\delta_{ai} \delta_{ap} \delta_{jk}
+\delta_{aj} \delta_{ak} \delta_{pi}
+\delta_{aj} \delta_{ap} \delta_{ik}
+\delta_{ak} \delta_{ap} \delta_{ij} \right)}{{(x_a-x_b)}^2}
\right.
\nonumber\\[2pt]
&&
\left.
+\sum_{a,b=1,a \ne b}^4 \frac{\left(
\delta_{ai} \delta_{bj} \delta_{kp}
+\delta_{ai} \delta_{bk} \delta_{pj}
+\delta_{ai} \delta_{bp} \delta_{jk}
+\delta_{aj} \delta_{bk} \delta_{pi}
+\delta_{aj} \delta_{bp} \delta_{ik}
+\delta_{ak} \delta_{bp} \delta_{ij} \right)}{{(x_a-x_b)}^2}
\right).
\nonumber
\eea
It is straightforward to verify that the Killing tensors are irreducible, i.e. they do not decompose into a combination of
the Killing vector fields (\ref{KV1}) and the metric (\ref{BM}). 

In a similar fashion, by increasing the number of particles 
in the original Calogero model, one can build
a $(d+2)$--dimensional Bohlin--type metric admitting Killing tensors 
of rank up to $d$.

\vspace{0.5cm}

\noindent
{\bf 5. Conclusion}\\

To summarize, inspired by the Bohlin transformation relating the 
planar harmonic oscillator to the Kepler problem, a variant of the
Eisenhart lift was proposed, in which a Lagrangian conservative dynamical system 
with $d$ degrees of freedom was embedded
into timelike geodesics of a conformally flat stationary metric defined on a $(d+2)$--dimensional 
space--time of the Lorentzian signature. The (anti) de Sitter space--time fitted nicely 
into the framework. 
The uplift was used to construct novel conformally flat metrics 
admitting higher rank Killing tensors. 

Let us recapitulate the main features of the construction. 
A potential of the original dynamical system 
determines the conformal factor of the metric, as well as it relates 
the coordinate time to the proper time.
Apart from Cartesian coordinates and a temporal 
variable, an extra spatial coordinate 
is needed in order to reproduce the conventional
nonrelativistic energy conservation condition within the general relativistic framework. 
In general, the metric admits
three Killing vector fields, 
which form the Lie algebra of the Poincar\'e group in $1+1$ dimensions. 
None of them is covariantly constant and, hence, the space--time does not belong to
the Kundt class. If a dynamical system at hand admits an integral of motion 
polynomial in momenta, it can be uplifted to
a Killing tensor. The uplift is different from that adopted within the conventional Eisenhart framework and
essentially relies upon the condition that the geodesic is timelike.

Turning to possible further developments, it would be interesting to use the 
Killing vector fields in 
(\ref{KV1}) to build the energy momentum tensor along the lines in \cite{FG} 
and study solutions to
the corresponding Einstein equation. 

A generalization to the case of time--dependent potentials is worth
exploring as well.

\vspace{0.5cm}

\noindent{\bf Acknowledgements}\\

\noindent
This work was supported by the RF Ministry of Science and Higher Education 
under the project FSWW-2026-0046.


\vspace{0.3cm}



\begin{thebibliography}{nn}

\bibitem{LE}
L. Eisenhart, {\it Dynamical trajectories and geodesics}, Ann. Math. {\bf 30} (1929) 591.
\bibitem{DBKP}
C. Duval, G. Burdet, H. K\"unzle, M. Perrin, {\it Bargmann structures and Newton--Cartan theory},
Phys. Rev. D {\bf 31} (1985) 1841.
\bibitem{FKK}
V. Frolov, P. Krtous, D. Kubiznak, {\it Black holes, hidden symmetries, and complete integrability}, 
Living Rev. Rel. {\bf 20} (2017) 6, arXiv:1705.05482.
\bibitem{DGH}
C. Duval, G.W. Gibbons, P.A. Horvathy, {\it Celestial mechanics, conformal structures and
gravitational waves}, Phys. Rev. D {\bf 43} (1991) 3907, arXiv:hep-th/0512188.
\bibitem{GHKW}
G.W. Gibbons, T. Houri, D. Kubiznak, C. Warnick, {\it Some spacetimes with higher rank Killing--Stackel tensors}, Phys. Lett. B {\bf 700} (2011) 68, arXiv:1103.5366.
\bibitem{GR}
G.W. Gibbons, C. Rugina, {\it  	
Goryachev--Chaplygin, Kovalevskaya, and Brdi\v{c}ka--Eardley--Nappi--Witten pp--waves spacetimes with higher rank St\"ackel--Killing tensors }, J. Math. Phys. {\bf 52} (2011) 122901, arXiv:1107.5987.
\bibitem{G}
A. Galajinsky, {\it Higher rank Killing tensors and Calogero model}, Phys. Rev. D {\bf 85} (2012) 085002, arXiv:1201.3085.
\bibitem{MCG}
M. Cariglia, G.W. Gibbons, {\it Generalised Eisenhart lift of the Toda chain}, J. Math. Phys. {\bf 55} (2014) 022701, arXiv:1312.2019.
\bibitem{CGHHKZ}
M. Cariglia, G.W. Gibbons, J.W. van Holten, P.A. Horv\'athy, P. Kosinski, 
P.M. Zhang, {\it Killing tensors and canonical geometry}, 
Class. Quant. Grav. {\bf 31} (2014) 125001, arXiv:1401.8195.
\bibitem{CGHHZ}
M. Cariglia, G.W. Gibbons, J.W. van Holten, P.A. Horvathy, P.M. Zhang, 
{\it Conformal Killing tensors and covariant Hamiltonian dynamics}, 
J. Math. Phys. {\bf 55} (2014) 122702, arxiv:1404.3422.
\bibitem{CG}
M. Cariglia, A. Galajinsky, {\it Ricci-flat spacetimes admitting higher rank Killing tensors}, Phys. Lett. B {\bf 744} (2015) 320, arXiv:1503.02162.
\bibitem{GF}
S. Filyukov, A. Galajinsky, {\it Self-dual metrics with maximally superintegrable geodesic flows}. Phys. Rev. D {\bf 91} (2015) 104020, arXiv:1504.03826.
\bibitem{CDGH}
 M. Cariglia, C. Duval, G.W. Gibbons, P.A. Horvathy, {\it Eisenhart lifts and symmetries of time-dependent systems}, 
 Annals Phys. {\bf 373} (2016) 631, arXiv:1605.01932.
\bibitem{GM}
A. Galajinsky, I. Masterov, {\it Eisenhart lift for higher derivative systems}, Phys. Lett. B {\bf 765} (2017) 86, arXiv:1611.04294.
\bibitem{FG} 	
A.P. Fordy, A. Galajinsky, {\it Eisenhart lift of 2--dimensional mechanics}, Eur. Phys. J. C {\bf 79} (2019) 301, arXiv:1901.03699.
\bibitem{BK}
K. Bartczak, P. Kosinski, {\it Herglotz's formalism, Eisenhart lift and 
Killing vectors}, arXiv:2508.21588.
\bibitem{G1}
A. Galajinsky, {\it Geometry of the isotropic oscillator driven by the conformal mode}, Eur. Phys. J. C {\bf 78} (2018) 72, arXiv:1712.00742.
\bibitem{CGGH}
M. Cariglia, A. Galajinsky, G.W. Gibbons, P.A. Horv\'athy, {\it Cosmological aspects of the 
Eisenhart--Duval lift}, Eur. Phys. J. C {\bf 78} (2018) 314, arXiv:1802.03370.
\bibitem{CH}
T. Chiba, T. Houri, {\it Eisenhart lift for scalar fields in the FLRW universe}, 
Class. Quant. Grav. {\bf 42} (2025) 055003, arxiv:2409.16325.
\bibitem{AP}
A. Paliathanasis, {\it Cosmological solutions in scalar–tensor theory via 
the Eisenhart–Duval lift}, Mod. Phys. Lett. A {\bf 40} (2025) 2550016, arxiv:2501.09356. 
\bibitem{B}
M.K. Bohlin,  {\it Note sur le probleme des deux corps et sur une
integration nouvelle dans le probleme des trois corps}, Bull Astrophysique
{\bf 28} (1911) 144.
\bibitem{S}
M.L. Saggio, {\it Bohlin transformation: the hidden
symmetry that connects Hooke to
Newton}, Eur. J. Phys. {\bf 34} (2013) 129.
\bibitem{AFF}
V. de Alfaro, S. Fubini, G. Furlan, {\it Conformal invariance in quantum mechanics}, Nuovo
Cim. A {\bf 34} (1976) 569.
\bibitem{Cal1}
F. Calogero, {\it Classical many--body problems amenable to exact treatments}, Lecture Notes
in Physics: Monographs {\bf 66}, Springer, 2001.
\bibitem{CGG}
S. Chanda, G.W. Gibbons, P. Guha, {\it Jacobi--Maupertuis metric and Kepler equation}, 
Int. J. Geom. Meth. Mod. Phys. {\bf 14} (2017) 1730002, arxiv:1612.07395.
\end{thebibliography}
\end{document}